\documentclass[11pt]{article}
\usepackage[margin=1in]{geometry}
\usepackage{amsmath,graphicx,url,bm}

\def\kslash{\rlap{\hspace{0.02cm}/}{k}}
\def\lslash{\rlap{\hspace{-0.02cm}/}{l}}

\def\beq{\begin{equation}}
\def\eeq#1{\label{#1}\end{equation}}
\def\eeqn{\end{equation}}

\def\beqa{\begin{eqnarray}}
\def\eeqa#1{\label{#1}\end{eqnarray}}
\def\eeqan{\end{eqnarray}}

\let\bar=\overbar

\def\Dslash{\not{\hbox{\kern-4pt $D$}}}
\def\dslash{\not{\hbox{\kern-2pt $\del$}}}

\def\msb{{\bar{\ssstyle M \kern -1pt S}}}

\def\Title#1{\begin{center} {\Large {\bf #1} } \end{center}}
\def\Author#1{\begin{center} {\normalsize {\sc #1} } \end{center}}
\def\Institution#1{\begin{center} {\normalsize {\it #1} } \end{center}}
\def\Abstract#1{\noindent {\normalsize {\bf Abstract:} {\normalfont #1}}}
\def\Conference{\vspace{4mm}\begin{raggedright} {\normalsize {\it Talk presented at the 2019 Meeting of the Division of Particles and Fields of the American Physical Society (DPF2019), July 29--August 2, 2019, Northeastern University, Boston, C1907293.} } \end{raggedright}\vspace{4mm}}

\begin{document}

%%%%%%%%%%%%%%%%%%%%%%%%%%%%%%%%%%%%%%%%%%%%%%%%%%%%%%%%%%%%%%%%%%%%%%%%%%%
%
% TITLE, AUTHOR, INSTITUTION, ABSTRACT ==> UPDATE
% 
%%%%%%%%%%%%%%%%%%%%%%%%%%%%%%%%%%%%%%%%%%%%%%%%%%%%%%%%%%%%%%%%%%%%%%%%%%%

\Title{The Proton Radius Puzzle}

\Author{Gil Paz}

\Institution{Department of Physics and Astronomy, \\
Wayne State University, Detroit, Michigan 48201, USA }

\Abstract{In 2010 the proton charge radius  was extracted for the first time from muonic hydrogen, a bound state of a muon and a proton. The value obtained was five standard deviations away from the regular hydrogen extraction. Taken at face value, this might be an indication of a new force in nature coupling to muons, but not to electrons. It also forces us to reexamine our understanding of the structure of the proton. Here I describe an ongoing theoretical research effort that seeks to address this ``proton radius puzzle". In particular, I will present the development of new effective field theoretical tools that seek to directly connect muonic hydrogen and muon-proton scattering.}

\Conference

%%%%%%%%%%%%%%%%%%%%%%%%%%%%%%%%%%%%%%%%%%%%%%%%%%%%%%%%%%%%%%%%%%%%%%%%%%%
%
% MAIN TEXT ==> UPDATE
% 
%%%%%%%%%%%%%%%%%%%%%%%%%%%%%%%%%%%%%%%%%%%%%%%%%%%%%%%%%%%%%%%%%%%%%%%%%%%

\section{Introduction}

How big is the proton? To answer such a question one needs to define how the proton size is measured. For example, one can use an electromagnetic probe to determine the proton's size. A ``one photon" electromagnetic interaction with an on-shell proton can be described by two form factors: $F_1$ and $F_2$. These form factors are functions of $q^2$, the square of the four-momentum transfer. Two different linear combinations of $F_1$ and $F_2$ define the ``electric" form factor: $G_E=F_1+q^2F_2/4M^2$, where $M$ is the proton mass, and the ``magnetic" form factor: $G_M=F_1+F_2$.  The slope of $G_E$ at $q^2=0$  \emph{defines} the proton charge radius $r_E^p$ via  $\left(r_E^p\right)^2 =6\,dG^{p}_E(q^2)/dq^2|_{q^2=0}$. Notice that this definition\footnote{In a specific frame, called the Breit frame, $r_E^p$ is often identified with the root-mean-square radius of the charge distribution, see \cite{Miller:2018ybm}  for a recent  discussion.} is Lorentz invariant \cite{Eides:2000xc,Eides:2007,Eides:2014swa}.    

Until 2010, the main method to extract $r_E^p$ was via electronic hydrogen spectroscopy. For example, the 2010 edition of the particle data book \cite{Nakamura:2010zzi} lists the value $r_E^p =0.8768(69)$ fm from the CODATA publication \cite{Mohr:2008fa}. Many extractions of $r_E^p$ from electron-proton scattering were listed in \cite{Nakamura:2010zzi} but they were not used. These span the period of 1963-2005 and the range of $0.8-0.9$ fm for $r_E^p$. In 2010, the first extraction of $r_E^p$ from \emph{muonic} hydrogen was reported as $r_E^p =0.84184(67)$ fm \cite{Pohl:2010zza}.  Surprisingly, it was five standard deviations lower than the regular hydrogen value. This nine-year old discrepancy is known as the ``proton radius puzzle" and it is still unresolved. The most recent value obtained  from muonic hydrogen is $r_E^p =0.84087(39)$ fm \cite{Antognini:1900ns}, while the most recent CODATA value is $r_E^p =0.8751(61)$ fm \cite{Mohr:2015ccw}.

The proton charge radius can be extracted in four types of experiments:  regular hydrogen spectroscopy, muonic hydrogen spectroscopy, electron-proton scattering, and muon-proton scattering. The proton radius puzzle has motivated new experiments in three of these areas.  For  muonic hydrogen spectroscopy there are no plans by other groups to repeat the measurement.  For electron-proton scattering a new  low-$Q^2$ electron-proton scattering experiment called ISR was performed by the A1 collaboration. They found $r_E^p =0.81(8)$ fm \cite{Mihovilovic:2016rkr} which cannot distinguish between the two values of $r_E^p$. Improved results by the same collaboration were presented at a 2018 MITP workshop on the proton radius puzzle \cite{PRP2018}. Another new low-$Q^2$ electron-proton scattering experiment called ``PRad" \cite{Gasparian:2017cgp} was recently performed at Jefferson Lab. Its results are not published yet\footnote{Preliminary results were presented at a conference \cite{DNP2018}, but are not available publicly.}.  Other planned scattering experiments were also presented at the MITP workshop \cite{PRP2018}. Muon-proton scattering is the least studied method to extract $r_E^p$. A new muon-proton scattering experiment called MUSE was built at the Paul Scherrer Institute \cite{Gilman:2017hdr}. It started taking data in 2019.  It is the first muon scattering measurement with the required precision to address the proton radius puzzle \cite{PRP2018}. For regular hydrogen spectroscopy several new measurements were published in the last two years with error bars comparable to the 2014 CODATA value. These are $r_E^p = 0.8335(95)$ fm  from $2S-4P$ transition \cite{Beyer:2017} by a group in Germany, $r_E^p =0.877(13)$ fm  from $1S-3S$ transition \cite{Fleurbaey:2018fih} by a group in France, and  $r_E^p =0.833(10)$ fm from $2S-2P$ transition \cite{Bezginov:2019mdi} by a group in Canada. A  preliminary result  from $1S-3S$ transition by the group in Germany was reported at the workshop  \cite{PRP2018}. They find $r_E^p$ smaller than that of \cite{Fleurbaey:2018fih}. Thus soon there will be two measurements of the \emph{same} $1S-3S$ transition that extract \emph{different} values of  $r_E^p$. 
     
A different method to extract $r_E^p$ is by using  lattice QCD. In the near future one can expect precise determinations of $r_E^p$ using lattice QCD that can distinguish between the conflicting experimental values \cite{Alexandrou}. 

In addition to the new experiments, there was also considerable activity on the theoretical side.  In the following I describe some advances in theory\footnote{The theory related to the extraction of $r_E^p$ from regular hydrogen spectroscopy is simpler compared to the other methods and no issues were raised about it.} related to electron-proton scattering (section \ref{sec:ep}), muonic hydrogen spectroscopy (section \ref{sec:Muonic}), and muon-proton scattering (section \ref{sec:mup}), focusing on work I was involved in. The conclusions are presented in section \ref{sec:conclusions}.  

\section{Advances in the theory of electron-proton scattering}\label{sec:ep}

Extractions of $r_E^p$ from  electron-proton cross section data or even the form factor itself require an extrapolation to $q^2=0$. Since $G_E$'s functional form is not known, such an extrapolation is not simple. Extractions that use \emph{different} functional forms for the \emph{same} data can lead to different values of $r_E^p$. The proliferation of functional forms might be one of the reasons that electron-proton scattering values of $r_E^p$ were not used in the 2010 PDG average \cite{Nakamura:2010zzi}. 

One of the important constraints on the form factor is its analytic structure. The form factor is analytic in the complex plane outside a cut that starts at the two-pion threshold\footnote{The $q^2$ threshold can be increased by including neutron and pion data \cite{Hill:2010yb, Epstein:2014zua,Paz:2011qr}. See \cite{Hoferichter:2015hva, Hoferichter:2016duk} for a recent analysis of the relevant pion data.} at $q^2=4m_\pi^2$ and extends to infinity. This implies, for example, that a simple Taylor expansion in $q^2$ cannot have a radius of convergence beyond $q^2=4m_\pi^2$. In order to incorporate the analytic structure constraints one can use the so-called $z$ expansion.  By changing variables from $q^2$ to $z$ we map the domain of analyticity onto the unit disk $|z|<1$. Since the form factor is analytic inside the unit $|z|=1$ circle it can be expanded as a Taylor series,
\begin{equation}
G_E^p(q^2) = \sum_{k=0}^\infty a_k \, z(q^2)^k.
\end{equation} 

For meson form factors the $z$ expansion is by now a standard default method. For example, the 2019 Flavor Lattice Averaging Group (FLAG) review  \cite{Aoki:2019cca} only shows $B$-meson form factors as a function of $z$ and not $q^2$. In \cite{Hill:2010yb} Richard Hill and I were the first to suggest to use the $z$ expansion for baryon form factors in general and for extraction of $r_E^p$ in particular. Since then it has been used to extract $r_E^p$ \cite{Hill:2010yb, Lee:2015jqa}, the proton magnetic radius \cite{Epstein:2014zua, Lee:2015jqa}, the neutron magnetic radius \cite{Epstein:2014zua}, the nucleon axial mass \cite{Bhattacharya:2011ah, Bhattacharya:2015mpa} and radius \cite{Meyer:2016oeg}, etc. It is also being used in lattice QCD studies of nucleon form factors.  

\begin{figure}[t]
\centering
\includegraphics[scale=0.65]{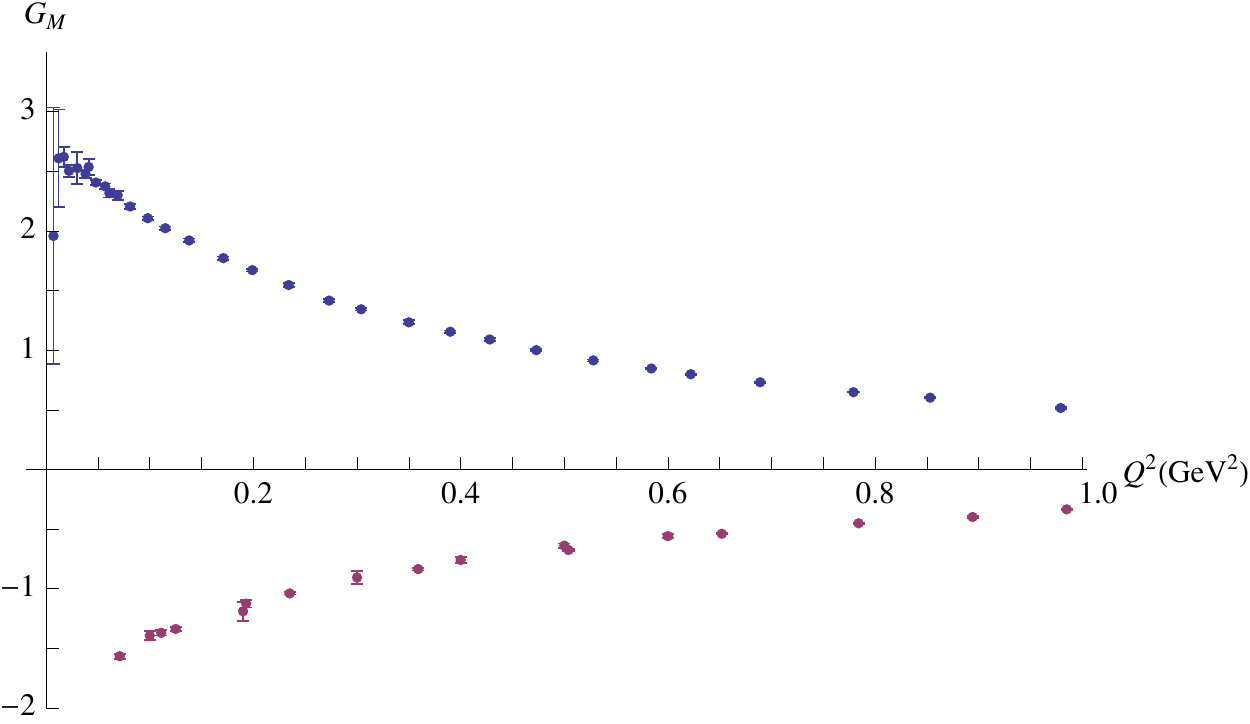}\quad
\includegraphics[scale=0.6]{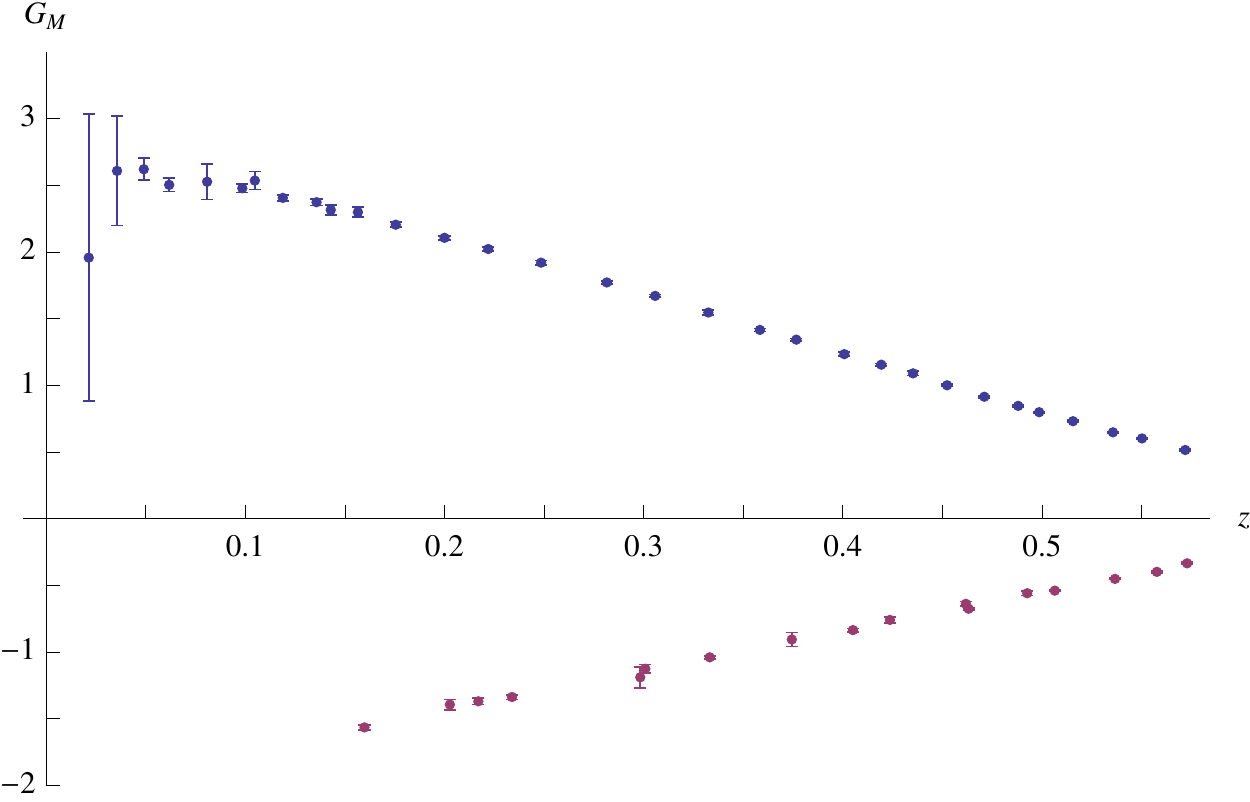}
\caption{Figure 2 of \cite{Epstein:2014zua}. Proton (above the horizontal axes) and neutron (below)  magnetic form factor data as a function of $Q^2$ (left) and as a function of $z$ (right). See \cite{Epstein:2014zua} for details. } 
\label{fig:GM}
\end{figure}

Here are two simple examples of the utility of the $z$ expansion. First, it implies that the historical dipole model of the form factor is not consistent with the analytic properties of the form factor.  Since $z^k$ are orthogonal over the unit $|z|=1$ circle, the coefficients $a_k$ are just Fourier coefficients, see \cite{Hill:2010yb}. A dipole form factor leads to a linear growth of $a_k$ with $k$ in contradiction to the analyticity of the form factor inside the unit circle. This is also in conflict with perturbative QCD \cite{Meyer:2016oeg}. Second,  form factor plotted as a function of $z$ can be simpler than the same data plotted as a function of $q^2$. See for example figure 2 of  \cite{Epstein:2014zua} reproduced here as figure \ref{fig:GM}. As a function of $Q^2=-q^2$  one would assume that there is a significant curvature, but in the $z$ variable the data is almost linear. This implies that in the $z$ variable one can extract with a reasonable uncertainty only an intercept and a slope (equivalent to $r_E^p$). To go beyond that requires better data. A similar phenomena is known from meson form factors  \cite{Hill:2006ub}.     
 
For an extraction of $r_E^p$ that does not depend on the number of terms in the series, the coefficients $a_k$ must be bounded. For lowest-lying meson form factors one can use constraints from unitarity. For baryon form factors unitarity only partially constrain the coefficients, see  \cite{Hill:2010yb} for details, and  one has to use other bounds as in  \cite{Hill:2010yb, Epstein:2014zua, Lee:2015jqa}. Some other $z$-expansion based studies do not bound the coefficients of the $z$ expansion \cite{Lorenz:2014vha, Lorenz:2014yda, Griffioen:2015hta} which may lead to issues in the extraction of $r_E^p$.

The values obtained using $z$-expansion analyses \cite{Hill:2010yb, Lee:2015jqa}  disfavor the muonic hydrogen result. More recently, lattice extractions of $r_E^p$ have used the $z$ expansion, see e.g.  \cite{Alexandrou:2017ypw,Jang:2018lup, Shintani:2018ozy}, although currently the errors are typically too large to distinguish between the two values of $r_E^p$.  

Besides the $z$-expansion analyses there were other recent extractions of $r_E^p$ that used other functional forms. These include  dipole \cite{Higinbotham:2015rja},  polynomial \cite{Griffioen:2015hta}, continued fraction \cite{Griffioen:2015hta}, and modified $z$ expansion \cite{Horbatsch:2015qda}. Some extractions include inputs from chiral effective field theory \cite{Horbatsch:2016ilr,Alarcon:2018zbz}. Most of these \cite{Higinbotham:2015rja,Griffioen:2015hta,Alarcon:2018zbz} favor the muonic hydrogen result. For pre-2010 extractions see  \cite{Nakamura:2010zzi}.

\section{Advances in the theory of muonic hydrogen spectroscopy}\label{sec:Muonic}

The first extraction of $r_E^p$ from muonic hydrogen has led to much discussion in the literature. It is reflected in the \emph{different} theoretical formula relating the measured energy level shift and $r_E^p$ used in 2010 in \cite{Pohl:2010zza} and in 2013 in \cite{Antognini:1900ns}. Due to its precision, the muonic hydrogen result involves a more complicated hadronic input, beyond a one-photon probe of the proton structure. In particular, as emphasized in \cite{Hill:2011wy}, two-photon effects are a potential source of uncertainty. The imaginary part of the two-photon exchange amplitude is related to experimental data: form factors and structure functions. Unfortunately the amplitude cannot be reconstructed from its imaginary part and the knowledge of a subtraction function $W_1(0,Q^2)$ is required. The subtraction function $W_1(0,Q^2)$ is not known exactly and unlike the imaginary part cannot be extracted directly from data. This introduces a potential large source of uncertainty.  Luckily, some information on $W_1(0,Q^2)$ can be obtained by considering its small and large $Q^2$ limits. 

Intuitively, in the small $Q^2$ limit the photon ``sees" the proton ``almost" like an elementary particle.  Non-Relativistic QED (NRQED) effective theory can be used \cite{Hill:2011wy} to give a rigorous interpretation to this intuition. Richard Hill and I used this to obtain the small $Q^2$ expansion \cite{Hill:2011wy}:  
\begin{equation}
W_1(0,Q^2) = 2a_p (2+a_p)  
  + 
  {Q^2\over m_p^2}
  \bigg\{ {2m_p^3 \bar{\beta} \over \alpha} -  a_p
  - \frac23\bigg[ (1+a_p)^2 m_p^2(r_M^p)^2 - m_p^2(r_E^p)^2 \bigg] \bigg\}\, , 
\end{equation}
where $a_p$ is the anomalous magnetic moment of the proton, $\bar{\beta}$ is the magnetic polarizability of the proton, and $r_M^p$ is the proton magnetic radius.  

Intuitively in the large $Q^2$ limit the photon ``sees" the quarks and gluons inside the proton. The large $Q^2$ expression for $W_1(0,Q^2)$  can be calculated using the operator product expansion.  There are two parts to the asymptotic form: spin-0 and spin-2.  The spin-0 contribution was calculated in 1978 by John Collins \cite{Collins:1978hi}. Richard Hill and I corrected it in 2016 \cite{Hill:2016bjv} and also calculated the spin-2 contribution which was unknown before.  All together the leading power result is \cite{Hill:2016bjv}
\begin{equation}
W_1(0,Q^2) = {2m_p^2\over Q^2} \bigg\{
  -\sum_f c_{1f} f_f^{(0)} + c_{1g}
  \tilde{f}_{g}^{(0)} 
  +\frac14 \bigg[ \sum_f \left( c_{2f} - c_{3f} \right) f^{(2)}_f
    + \left( c_{2g} - c_{3g} \right) f^{(2)}_g \bigg]
    \bigg\} \,, 
\end{equation}
where $c_{if}$ and $c_{ig}$ with $i=1,2,3$ are the Wilson coefficients, $f_f^{(0)}$ and  $\tilde{f}_{g}^{(0)}$ are the spin-0 matrix elements,  and $f^{(2)}_f$ and $f^{(2)}_g$ are the spin-2 matrix elements.  See \cite{Hill:2016bjv} for details.

In \cite{Hill:2016bjv} we interpolated the two limits to give an estimate for the contribution of $W_1(0,Q^2)$ to two-photon exchange effects. The uncertainty on the interpolation is larger than in \cite{Antognini:1900ns}, but it is too small to explain the discrepancy. The estimate in \cite{Hill:2016bjv} is consistent with the literature \cite{Pachucki:1999zza, Martynenko:2005rc, Nevado:2007dd, Carlson:2011zd, Birse:2012eb, Gorchtein:2013yga, Alarcon:2013cba, Peset:2014jxa}. On the other hand, \cite{Miller:2012ne} finds a much larger uncertainty. Ultimately one would like to probe the muon-proton two-photon exchange effects by using a different method such as muon-proton scattering.

\section{Advances in the theory of muon-proton scattering} \label{sec:mup} 
The MUSE experiment  is intended to extract $r_E^p$ for the first time from muon-proton scattering . It can also be sensitive to possibly anomalous two-photon exchange effects.  In making predictions for MUSE, a phenomenological approach was taken in a series of papers by Oleksandr Tomalak and Marc Vanderhaeghen \cite{Tomalak:2014dja, Tomalak:2015hva, Tomalak:2017owk, Tomalak:2018jak}. My collaborators and I have studied the utility of effective field theory (EFT) methods. 

In muonic hydrogen the muon's typical momentum is $m\alpha\sim1$ MeV, and both the muon and the proton can be treated non-relativistically.  For MUSE the muon's typical momentum is about the muon mass $m\sim100$ MeV, and the muon must be treated relativistically, while the proton can be treated non-relativistically. Richard Hill, Gabriel Lee, Mikhail Solon and I suggested an EFT, called QED-NRQED, that is applicable for such kinematics\footnote{The dynamical degrees of freedom of this theory are proton, muon, and photon. The pion is not included as a dynamical degree of freedom. This is different from an earlier EFT applicable to the MUSE kinematics considered by Antonio Pineda in \cite{Pineda:2002as, Pineda:2004mx} that contains very similar operators. } in \cite{Hill:2012rh}.

Steven Dye, Matthew Gonderinger, and I  studied some aspects of this EFT in \cite{Dye:2016uep}. Denoting by $m (M)$ the muon (proton) mass and using  $Z=1$ for a proton, we showed that one-photon exchange ${\cal O} (Z\alpha)$ QED-NRQED scattering at power $1/M^2$ reproduces Rosenbluth scattering \cite{Rosenbluth:1950yq}, and the two-photon exchange ${\cal O} (Z^2\alpha^2)$ QED-NRQED scattering at leading power reproduces the scattering of a relativistic fermion off a static potential \cite{Dalitz:1951ah, Itzykson:1980rh}.  

Two photon exchange effects start at ${\cal O} (Z^2\alpha^2)$ and power $1/M^2$. For QED-NRQED they appear as two four-fermion operators:
%, one spin-independent and one spin-dependent:
\begin{equation}\label{contactQN} 
{\cal L}_{\psi\ell}=\dfrac{b_1}{M^2}\psi^\dagger\psi\,\bar \ell\gamma^0\ell+\dfrac{b_2}{M^2}\psi^\dagger\sigma^i\psi\,\bar \ell\gamma^i\gamma^5\ell +{\cal O}\left(1/M^3\right),
\end{equation}

In a recent paper \cite{Dye:2018rgg} Steven Dye, Matthew Gonderinger, and I determined  $b_1$ and $b_2$ at  ${\cal O}(Z^2\alpha^2)$. For that we calculated the $\ell+p\to\ell+p$ \emph{off-shell}  forward scattering amplitude at ${\cal O}(Z^2\alpha^2)$ and power $1/M^2$ in the effective and full theory  in both Feynman and Coulomb gauges. We considered two cases of full theories:  a toy example of a non-relativistic point particle, and the real proton which is described by a hadronic tensor.

For the toy example we found $b_1^{\mbox{\scriptsize p.p.}}=0$ and $b_2^{\mbox{\scriptsize p.p.}}=Q_l^2Z^2\alpha^2\left[{16}/{3}+\log\left({M}/{2\Lambda}\right)\right]$, where $\Lambda$ is the UV cutoff of QED-NRQED. \emph{Surprisingly} $b_1^{\mbox{\scriptsize p.p.}}=0$ at ${\cal O}(Z^2\alpha^2)$. For the case of the real proton we give implicit expressions for $b_1$ and $b_2$ in terms of the components of the hadronic tensor.  Considering only the contribution of $F_1(0)$, $F_2(0)$ and $M^2 F_1^\prime(0)$ to the Wilson coefficients we found: $b_1(\alpha^2Q_\ell^2)^{-1}=0+\cdots,\, b_2(\alpha^2Q_\ell^2)^{-1}=F_1(0)^2\left[{16}/{3}+\log\left({M}/{2\Lambda}\right)\right]+F_1(0)F_2(0) {16}/{3}+F_2(0)^2\left[{17}/{12}-\log\left({M}/{2\Lambda}\right)+3\log\left({Q}/{M}\right)\right]/2+\cdots\,$. The ellipsis denotes  $\mbox{non } F_1(0), F_2(0), M^2 F_1^\prime(0)$ terms. See \cite{Dye:2018rgg} for details.  Surprisingly, \emph{again} there is no contribution to $b_1$. Why?

The vanishing of $b_1$  at ${\cal O}(Z^2\alpha^2)$ arises from a combination of two phenomena. On the EFT side the diagrams involve the propagator $(\pm l^0-{\vec l^{\,\,2}}/{2M}+i\epsilon)^{-1}$, with the plus (minus) sign corresponds to a direct (crossed) diagram. Expanding in $1/M$, the $1/M^2$ terms come with opposite signs.  Direct and crossed diagrams usually appear as a sum for spin-independent terms and cancel each other. The exception are the IR divergent terms, but these must cancel in the matching.  

On the full theory side the amplitude involves integrals over the hadronic tensor $W^{\mu\nu}(p,l)$. Defining $k=(m,\vec 0)$ and taking the limit $m\to 0\Rightarrow k\to 0$ the full theory amplitude is 
\begin{equation}
\dfrac{i{\cal M}_{\mbox{\scriptsize Full}}}{-Q_\ell^2e^4}
=\int\dfrac{d^4l}{(2\pi)^4}\dfrac{\bar u\gamma_\mu(\kslash-\lslash+m)\gamma_\nu u}{(k-l)^2-m^2}\dfrac{W^{\mu\nu}(p,l) }{\left(l^2-\lambda^2\right)^2}\to \int\dfrac{d^4l}{(2\pi)^4}\dfrac{\bar u\gamma_\mu(-\lslash)\gamma_\nu u}{l^2}\dfrac{W^{\mu\nu}(p,l) }{\left(l^2-\lambda^2\right)^2}\,.
\end{equation}
Translation invariance implies $W^{\mu\nu}(p,l)=W^{\nu\mu}(p,-l)$ \cite{Dye:2018rgg}. Since the full theory spin-independent amplitude is symmetric in $\mu\leftrightarrow\nu$, it  vanishes for $m\to 0$.

The combination of the two phenomena leads to the vanishing of $b_1$  at ${\cal O}(Z^2\alpha^2)$. Notice that it does not obviously follow from a symmetry of the EFT. It might be a one-loop or power $1/M^2$ ``accident".  In  \cite{Dye:2018rgg} we showed that for the toy example of a point-particle full theory there is a term that contributes to spin-independent matching coefficient at power $1/M^3$.  

Beyond the proton radius puzzle itself, this result can be of interest in physics beyond the standard model, where generating hierarchies, even ``little" ones, between the weak scale and the scale of new physics is an active topic of research. It would be interesting to see if the vanishing of $b_1$ at ${\cal O}(Z^2\alpha^2)$ can be used  to generate such hierarchies.

\section{Conclusions}\label{sec:conclusions}
The proton radius puzzle has motivated the reevaluation of our understanding of the proton. It has led to new experiments and theoretical advances. Here I described three such advances in theory that I was involved in: electron-proton scattering, muonic hydrogen spectroscopy, and muon-proton scattering. 

For the theory of electron-proton scattering one of the important advances is the introduction of the $z$ expansion for baryon form factors in \cite{Hill:2010yb}. Based on its previous success in describing meson form factors, one can expect this method to become more and more prevalent in extraction of $r_E^p$ from electron-proton scattering. Extractions based on the $z$ expansion generally disfavor the muonic hydrogen result. Even if the regular and muonic hydrogen values were to agree, this is an issue that will need to be resolved. See \cite{Lee:2015jqa} for further discussion.

For the theory of muonic hydrogen spectroscopy one of the challenges is the calculation of two-photon exchange effects. The fact that we cannot reproduce the full hadronic tensor from experimental data implies that there is an inherent uncertainty that cannot be reduced. The missing piece is the subtraction function $W_1(0,Q^2)$. Its small $Q^2$ expansion is known for some time.  In 2016 its large $Q^2$ was calculated for the first time \cite{Hill:2016bjv} combining the spin-2 part and correcting the spin-0 part calculated in 1978 \cite{Collins:1978hi}.  The high and low $Q^2$ constraints should be fulfilled by any theory that aims to estimate the two-photon exchange effects. A simple interpolation of the two limits done in  \cite{Hill:2016bjv} finds a larger uncertainty than the one used in analyzing experimental data, but not large enough to explain the discrepancy.

For the theory of muon-proton scattering one of the main tasks is making prediction for MUSE, the new muon-proton scattering experiment. As for spectroscopy, the main challenge are two-photon exchange effects. My collaborators and I have suggested to use an EFT \cite{Hill:2012rh}, called QED-NRQED, to calculate muon-proton scattering. The Wilson coefficients of the effective theory up to dimension six depended on the proton's charge, magnetic moment, $r_E^p$, and two types of two-photon exchange effects: spin-independent and spin-dependent. In \cite{Dye:2016uep} we studied one-photon exchange effects up to sub-sub-leading power and two-photon exchange effects at leading power and showed that they reproduce known results from the literature. Recently we have calculated the Wilson coefficients of the two contact interactions in terms of the components of the hadronic tensor. Surprisingly, the spin-independent Wilson coefficient vanishes at  ${\cal O}(Z^2\alpha^2)$. This does not follow from any symmetry of the effective theory. It also implies that MUSE will be much less sensitive to such effects, but its extraction of $r_E^p$ will be more robust. 

In summary, after more than nine years of experimental and theoretical work we have learned a lot but the proton radius puzzle is still puzzling... .

\section*{Acknowledgements}
This work was supported by the U.S. Department of Energy grant DE-SC0007983 and by a Career Development Chair award from Wayne State University. The work reported here was supported by grants from DOE, NSF, NIST, Simons Foundation, and Fermilab.


\begin{thebibliography}{99}

\bibitem{Eides:2000xc} 
  M.~I.~Eides, H.~Grotch and V.~A.~Shelyuto,
  %``Theory of light hydrogen - like atoms,''
  Phys.\ Rept.\  {\bf 342}, 63 (2001)
  %doi:10.1016/S0370-1573(00)00077-6
  [hep-ph/0002158].
  %%CITATION = doi:10.1016/S0370-1573(00)00077-6;%%
  
\bibitem{Eides:2007} 
M.~I.~Eides, H.~Grotch, and V.~A.~Shelyuto, ``Theory of Light Hydrogenic Bound States," Springer 2007, Berlin,
Heidelberg, New York.
  
  
\bibitem{Eides:2014swa} 
  M.~I.~Eides,
  ``On Some Recent Ideas on the Proton Radius Puzzle and Lepton Anomalous Magnetic Moments,''
  %Phys.\ Rev.\ D {\bf 90}, no. 5, 057301 (2014)
 % doi:10.1103/PhysRevD.90.057301
  [arXiv:1402.5860 [hep-ph]].
  %%CITATION = doi:10.1103/PhysRevD.90.057301;%%  


\bibitem{Miller:2018ybm} 
  G.~A.~Miller,
  %``Defining the proton radius: A unified treatment,''
  Phys.\ Rev.\ C {\bf 99}, no. 3, 035202 (2019)
 % doi:10.1103/PhysRevC.99.035202
  [arXiv:1812.02714 [nucl-th]].
  %%CITATION = doi:10.1103/PhysRevC.99.035202;%%
  
\bibitem{Nakamura:2010zzi} 
  K.~Nakamura {\it et al.} [Particle Data Group],
  %``Review of particle physics,''
  J.\ Phys.\ G {\bf 37}, 075021 (2010).
  %doi:10.1088/0954-3899/37/7A/075021
  %%CITATION = doi:10.1088/0954-3899/37/7A/075021;%%  
  
\bibitem{Mohr:2008fa} 
  P.~J.~Mohr, B.~N.~Taylor and D.~B.~Newell,
  %``CODATA Recommended Values of the Fundamental Physical Constants: 2006,''
  Rev.\ Mod.\ Phys.\  {\bf 80}, 633 (2008)
  %doi:10.1103/RevModPhys.80.633
  [arXiv:0801.0028 [physics.atom-ph]].
  %%CITATION = doi:10.1103/RevModPhys.80.633;%%
  

\bibitem{Pohl:2010zza} 
  R.~Pohl {\it et al.},
  %``The size of the proton,''
  Nature {\bf 466}, 213 (2010).
  %doi:10.1038/nature09250
  %%CITATION = doi:10.1038/nature09250;%%
  
\bibitem{Antognini:1900ns} 
  A.~Antognini {\it et al.},
  %``Proton Structure from the Measurement of $2S-2P$ Transition Frequencies of Muonic Hydrogen,''
  Science {\bf 339}, 417 (2013).
  %doi:10.1126/science.1230016
  %%CITATION = doi:10.1126/science.1230016;%%
  
\bibitem{Mohr:2015ccw} 
  P.~J.~Mohr, D.~B.~Newell and B.~N.~Taylor,
  %``CODATA Recommended Values of the Fundamental Physical Constants: 2014,''
  Rev.\ Mod.\ Phys.\  {\bf 88},  035009 (2016)
  %doi:10.1103/RevModPhys.88.035009
  [arXiv:1507.07956 [physics.atom-ph]].
  %%CITATION = doi:10.1103/RevModPhys.88.035009;%%    
  
\bibitem{Mihovilovic:2016rkr} 
  M.~Mihovilovi\v{c} {\it et al.},
  %``First measurement of proton's charge form factor at very low $Q^2$ with initial state radiation,''
  Phys.\ Lett.\ B {\bf 771}, 194 (2017)
  %doi:10.1016/j.physletb.2017.05.031
  [arXiv:1612.06707 [nucl-ex]].
  %%CITATION = doi:10.1016/j.physletb.2017.05.031;%%  
  
\bibitem{PRP2018}
See the talks at the 2018 MITP workshop ``Precision Measurements and Fundamental Physics: The Proton Radius Puzzle and Beyond"\\
\url{https://indico.mitp.uni-mainz.de/event/132/timetable/#all}    

  
  \bibitem{Gasparian:2017cgp} 
  A.~H.~Gasparian [PRad Collaboration],
  %``The New Proton Radius Experiment at Jefferson Lab,''
  JPS Conf.\ Proc.\  {\bf 13}, 020052 (2017).
  %doi:10.7566/JPSCP.13.020052
  %%CITATION = doi:10.7566/JPSCP.13.020052;%%    
  
  \bibitem{DNP2018}
\url{http://meetings.aps.org/Meeting/HAW18/APS_epitome}

\bibitem{Gilman:2017hdr} 
  R.~Gilman {\it et al.} [MUSE Collaboration],
  %``Technical Design Report for the Paul Scherrer Institute Experiment R-12-01.1: Studying the Proton "Radius" Puzzle with $\mu p$ Elastic Scattering,''
  arXiv:1709.09753 [physics.ins-det].
  %%CITATION = ARXIV:1709.09753;%%  

  
  
\bibitem{Beyer:2017}
A. Beyer {\it et al.}, 
Science {\bf 358}, 79 (2017).

\bibitem{Fleurbaey:2018fih} 
  H.~Fleurbaey {\it et al.},
  %``New Measurement of the $1S-3S$ Transition Frequency of Hydrogen: Contribution to the Proton Charge Radius Puzzle,''
  Phys.\ Rev.\ Lett.\  {\bf 120},  183001 (2018)
  %doi:10.1103/PhysRevLett.120.183001
  [arXiv:1801.08816 [physics.atom-ph]].
  %%CITATION = doi:10.1103/PhysRevLett.120.183001;%%
  
\bibitem{Bezginov:2019mdi} 
  N.~Bezginov, T.~Valdez, M.~Horbatsch, A.~Marsman, A.~C.~Vutha and E.~A.~Hessels,
  %``A measurement of the atomic hydrogen Lamb shift and the proton charge radius,''
  Science {\bf 365},  1007 (2019).
 % doi:10.1126/science.aau7807
  %%CITATION = doi:10.1126/science.aau7807;%%
 
\bibitem{Alexandrou} Constantia Alexandrou \emph{private communication}. 

\bibitem{Hill:2010yb} 
  R.~J.~Hill and G.~Paz,
  %``Model independent extraction of the proton charge radius from electron scattering,''
  Phys.\ Rev.\ D {\bf 82}, 113005 (2010)
  %doi:10.1103/PhysRevD.82.113005
  [arXiv:1008.4619 [hep-ph]].
  %%CITATION = doi:10.1103/PhysRevD.82.113005;%%  
  
\bibitem{Epstein:2014zua} 
  Z.~Epstein, G.~Paz and J.~Roy,
  %``Model independent extraction of the proton magnetic radius from electron scattering,''
  Phys.\ Rev.\ D {\bf 90},  074027 (2014)
 % doi:10.1103/PhysRevD.90.074027
  [arXiv:1407.5683 [hep-ph]].
  %%CITATION = doi:10.1103/PhysRevD.90.074027;%%  
  
\bibitem{Paz:2011qr} 
  G.~Paz,
  %``The Charge Radius of the Proton,''
  AIP Conf.\ Proc.\  {\bf 1441}, no. 1, 146 (2012)
  %doi:10.1063/1.3700495
  [arXiv:1109.5708 [hep-ph]].
  %%CITATION = doi:10.1063/1.3700495;%%  
  
\bibitem{Hoferichter:2015hva} 
  M.~Hoferichter, J.~Ruiz de Elvira, B.~Kubis and U.~G.~Mei§ner,
  %``RoyÐSteiner-equation analysis of pionÐnucleon scattering,''
  Phys.\ Rept.\  {\bf 625}, 1 (2016)
  %doi:10.1016/j.physrep.2016.02.002
  [arXiv:1510.06039 [hep-ph]].
  %%CITATION = doi:10.1016/j.physrep.2016.02.002;%%  
  
  
\bibitem{Hoferichter:2016duk} 
  M.~Hoferichter, B.~Kubis, J.~Ruiz de Elvira, H.-W.~Hammer and U.-G.~Mei§ner,
  %``On the $\pi\pi$ continuum in the nucleon form factors and the proton radius puzzle,''
  Eur.\ Phys.\ J.\ A {\bf 52}, no. 11, 331 (2016)
  %doi:10.1140/epja/i2016-16331-7
  [arXiv:1609.06722 [hep-ph]].
  %%CITATION = doi:10.1140/epja/i2016-16331-7;%%  
   
\bibitem{Aoki:2019cca} 
  S.~Aoki {\it et al.} [Flavour Lattice Averaging Group],
  %``FLAG Review 2019,''
  arXiv:1902.08191 [hep-lat].
  %%CITATION = ARXIV:1902.08191;%%  

\bibitem{Lee:2015jqa} 
  G.~Lee, J.~R.~Arrington and R.~J.~Hill,
  %``Extraction of the proton radius from electron-proton scattering data,''
  Phys.\ Rev.\ D {\bf 92}, 013013 (2015)
  %doi:10.1103/PhysRevD.92.013013
  [arXiv:1505.01489 [hep-ph]].
  %%CITATION = doi:10.1103/PhysRevD.92.013013;%%  
 
\bibitem{Bhattacharya:2011ah} 
  B.~Bhattacharya, R.~J.~Hill and G.~Paz,
  %``Model independent determination of the axial mass parameter in quasielastic neutrino-nucleon scattering,''
  Phys.\ Rev.\ D {\bf 84}, 073006 (2011)
  %doi:10.1103/PhysRevD.84.073006
  [arXiv:1108.0423 [hep-ph]].
  %%CITATION = doi:10.1103/PhysRevD.84.073006;%%  
  
\bibitem{Bhattacharya:2015mpa} 
  B.~Bhattacharya, G.~Paz and A.~J.~Tropiano,
  %``Model-independent determination of the axial mass parameter in quasielastic antineutrino-nucleon scattering,''
  Phys.\ Rev.\ D {\bf 92}, no. 11, 113011 (2015)
  %doi:10.1103/PhysRevD.92.113011
  [arXiv:1510.05652 [hep-ph]].
  %%CITATION = doi:10.1103/PhysRevD.92.113011;%%        
  
\bibitem{Meyer:2016oeg} 
  A.~S.~Meyer, M.~Betancourt, R.~Gran and R.~J.~Hill,
  %``Deuterium target data for precision neutrino-nucleus cross sections,''
  Phys.\ Rev.\ D {\bf 93}, no. 11, 113015 (2016)
  %doi:10.1103/PhysRevD.93.113015
  [arXiv:1603.03048 [hep-ph]].
  %%CITATION = doi:10.1103/PhysRevD.93.113015;%%  
  
\bibitem{Hill:2006ub} 
  R.~J.~Hill,
  %``The Modern description of semileptonic meson form factors,''
  eConf C {\bf 060409}, 027 (2006)
  [hep-ph/0606023].
  %%CITATION = HEP-PH/0606023;%%
  
\bibitem{Lorenz:2014vha} 
  I.~T.~Lorenz and U.~G.~Mei§ner,
  %``Reduction of the proton radius discrepancy by 3?,''
  Phys.\ Lett.\ B {\bf 737}, 57 (2014)
  %doi:10.1016/j.physletb.2014.08.010
  [arXiv:1406.2962 [hep-ph]].
  %%CITATION = doi:10.1016/j.physletb.2014.08.010;%%  
  
\bibitem{Lorenz:2014yda} 
  I.~T.~Lorenz, U.~G.~Mei§ner, H.-W.~Hammer and Y.-B.~Dong,
  %``Theoretical Constraints and Systematic Effects in the Determination of the Proton Form Factors,''
  Phys.\ Rev.\ D {\bf 91}, no. 1, 014023 (2015)
  %doi:10.1103/PhysRevD.91.014023
  [arXiv:1411.1704 [hep-ph]].
  %%CITATION = doi:10.1103/PhysRevD.91.014023;%%  
  
\bibitem{Griffioen:2015hta} 
  K.~Griffioen, C.~Carlson and S.~Maddox,
  %``Consistency of electron scattering data with a small proton radius,''
  Phys.\ Rev.\ C {\bf 93}, no. 6, 065207 (2016)
 % doi:10.1103/PhysRevC.93.065207
  [arXiv:1509.06676 [nucl-ex]].
  %%CITATION = doi:10.1103/PhysRevC.93.065207;%%  
  
\bibitem{Alexandrou:2017ypw} 
  C.~Alexandrou, M.~Constantinou, K.~Hadjiyiannakou, K.~Jansen, C.~Kallidonis, G.~Koutsou and A.~Vaquero Aviles-Casco,
  %``Nucleon electromagnetic form factors using lattice simulations at the physical point,''
  Phys.\ Rev.\ D {\bf 96}, no. 3, 034503 (2017)
  %doi:10.1103/PhysRevD.96.034503
  [arXiv:1706.00469 [hep-lat]].
  %%CITATION = doi:10.1103/PhysRevD.96.034503;%%  
  
  
\bibitem{Jang:2018lup} 
  Y.~C.~Jang, T.~Bhattacharya, R.~Gupta, H.~W.~Lin and B.~Yoon,
  %``Nucleon Axial and Electromagnetic Form Factors,''
  EPJ Web Conf.\  {\bf 175}, 06033 (2018)
  %doi:10.1051/epjconf/201817506033
  [arXiv:1801.01635 [hep-lat]].
  %%CITATION = doi:10.1051/epjconf/201817506033;%%  
  

\bibitem{Shintani:2018ozy} 
  E.~Shintani, K.~I.~Ishikawa, Y.~Kuramashi, S.~Sasaki and T.~Yamazaki,
  %``Nucleon form factors and root-mean-square radii on a (10.8 fm$)^4$ lattice at the physical point,''
  arXiv:1811.07292 [hep-lat].
  %%CITATION = ARXIV:1811.07292;%%  

\bibitem{Higinbotham:2015rja} 
  D.~W.~Higinbotham, A.~A.~Kabir, V.~Lin, D.~Meekins, B.~Norum and B.~Sawatzky,
  %``Proton radius from electron scattering data,''
  Phys.\ Rev.\ C {\bf 93},  055207 (2016)
  %doi:10.1103/PhysRevC.93.055207
  [arXiv:1510.01293 [nucl-ex]].
  %%CITATION = doi:10.1103/PhysRevC.93.055207;%%  
  
  
\bibitem{Horbatsch:2015qda} 
  M.~Horbatsch and E.~A.~Hessels,
  %``Evaluation of the strength of electron-proton scattering data for determining the proton charge radius,''
  Phys.\ Rev.\ C {\bf 93},  015204 (2016)
  %doi:10.1103/PhysRevC.93.015204
  [arXiv:1509.05644 [nucl-ex]].
  %%CITATION = doi:10.1103/PhysRevC.93.015204;%%  
    
\bibitem{Horbatsch:2016ilr} 
  M.~Horbatsch, E.~A.~Hessels and A.~Pineda,
  %``Proton radius from electron-proton scattering and chiral perturbation theory,''
  Phys.\ Rev.\ C {\bf 95}, 035203 (2017)
  %doi:10.1103/PhysRevC.95.035203
  [arXiv:1610.09760 [nucl-th]].
  %%CITATION = doi:10.1103/PhysRevC.95.035203;%%  
  
\bibitem{Alarcon:2018zbz} 
  J.~M.~Alarc\'on, D.~W.~Higinbotham, C.~Weiss and Z.~Ye,
  %``Proton charge radius extraction from electron scattering data using dispersively improved chiral effective field theory,''
  Phys.\ Rev.\ C {\bf 99}, 044303 (2019)
  %doi:10.1103/PhysRevC.99.044303
  [arXiv:1809.06373 [hep-ph]].
  %%CITATION = doi:10.1103/PhysRevC.99.044303;%%  

\bibitem{Hill:2011wy} 
  R.~J.~Hill and G.~Paz,
  %``Model independent analysis of proton structure for hydrogenic bound states,''
  Phys.\ Rev.\ Lett.\  {\bf 107}, 160402 (2011)
  %doi:10.1103/PhysRevLett.107.160402
  [arXiv:1103.4617 [hep-ph]].
  %%CITATION = doi:10.1103/PhysRevLett.107.160402;%%    
  
\bibitem{Collins:1978hi} 
  J.~C.~Collins,
  %``Renormalization of the Cottingham Formula,''
  Nucl.\ Phys.\ B {\bf 149}, 90 (1979)
  Erratum: [Nucl.\ Phys.\ B {\bf 153}, 546 (1979)].
 % doi:10.1016/0550-3213(79)90158-5
  %%CITATION = doi:10.1016/0550-3213(79)90158-5;%%      
  
\bibitem{Hill:2016bjv} 
  R.~J.~Hill and G.~Paz,
  %``Nucleon spin-averaged forward virtual Compton tensor at large $Q^2$,''
  Phys.\ Rev.\ D {\bf 95},  094017 (2017)
  %doi:10.1103/PhysRevD.95.094017
  [arXiv:1611.09917 [hep-ph]].
  %%CITATION = doi:10.1103/PhysRevD.95.094017;%%  
  
   
  
\bibitem{Pachucki:1999zza} 
  K.~Pachucki,
  %``Proton structure effects in muonic hydrogen,''
  Phys.\ Rev.\ A {\bf 60}, 3593 (1999)
%  doi:10.1103/PhysRevA.60.3593
  %%CITATION = doi:10.1103/PhysRevA.60.3593;%%  
  
  
\bibitem{Martynenko:2005rc} 
  A.~P.~Martynenko,
  %``Proton polarizability effect in the Lamb shift of the hydrogen atom,''
  Phys.\ Atom.\ Nucl.\  {\bf 69}, 1309 (2006).
%  doi:10.1134/S1063778806080072
  [hep-ph/0509236].
  %%CITATION = doi:10.1134/S1063778806080072;%%
  
 

\bibitem{Nevado:2007dd} 
  D.~Nevado and A.~Pineda,
  %``Forward virtual Compton scattering and the Lamb shift in chiral perturbation theory,''
  Phys.\ Rev.\ C {\bf 77}, 035202 (2008)
%  doi:10.1103/PhysRevC.77.035202
  [arXiv:0712.1294 [hep-ph]].
  %%CITATION = doi:10.1103/PhysRevC.77.035202;%%
  
  
  
\bibitem{Carlson:2011zd} 
  C.~E.~Carlson and M.~Vanderhaeghen,
  %``Higher order proton structure corrections to the Lamb shift in muonic hydrogen,''
  Phys.\ Rev.\ A {\bf 84}, 020102 (2011).
%  doi:10.1103/PhysRevA.84.020102
  [arXiv:1101.5965 [hep-ph]].
  %%CITATION = doi:10.1103/PhysRevA.84.020102;%%  
  
\bibitem{Birse:2012eb} 
  M.~C.~Birse and J.~A.~McGovern,
  %``Proton polarisability contribution to the Lamb shift in muonic hydrogen at fourth order in chiral perturbation theory,''
  Eur.\ Phys.\ J.\ A {\bf 48}, 120 (2012)
%  doi:10.1140/epja/i2012-12120-8
%  [arXiv:1206.3030 [hep-ph]].
  %%CITATION = doi:10.1140/epja/i2012-12120-8;%%
  
 
\bibitem{Gorchtein:2013yga} 
  M.~Gorchtein, F.~J.~Llanes-Estrada and A.~P.~Szczepaniak,
  %``Muonic-hydrogen Lamb shift: Dispersing the nucleon-excitation uncertainty with a finite-energy sum rule,''
  Phys.\ Rev.\ A {\bf 87},  052501 (2013)
%  doi:10.1103/PhysRevA.87.052501
 [arXiv:1302.2807 [nucl-th]].
  %%CITATION = doi:10.1103/PhysRevA.87.052501;%%
  
  

\bibitem{Alarcon:2013cba} 
  J.~M.~Alarc\'on, V.~Lensky and V.~Pascalutsa,
  %``Chiral perturbation theory of muonic hydrogen Lamb shift: polarizability contribution,''
  Eur.\ Phys.\ J.\ C {\bf 74},  2852 (2014)
%  doi:10.1140/epjc/s10052-014-2852-0
 [arXiv:1312.1219 [hep-ph]].
  %%CITATION = doi:10.1140/epjc/s10052-014-2852-0;%% 
  

  
\bibitem{Peset:2014jxa} 
  C.~Peset and A.~Pineda,
  %``The two-photon exchange contribution to muonic hydrogen from chiral perturbation theory,''
  Nucl.\ Phys.\ B {\bf 887}, 69 (2014)
%  doi:10.1016/j.nuclphysb.2014.07.027
  [arXiv:1406.4524 [hep-ph]].
  %%CITATION = doi:10.1016/j.nuclphysb.2014.07.027;%%  
  
  

\bibitem{Miller:2012ne} 
  G.~A.~Miller,
  %``Proton Polarizability Contribution: Muonic Hydrogen Lamb Shift and Elastic Scattering,''
  Phys.\ Lett.\ B {\bf 718}, 1078 (2013)
  %doi:10.1016/j.physletb.2012.11.016
  [arXiv:1209.4667 [nucl-th]].
  %%CITATION = doi:10.1016/j.physletb.2012.11.016;%%  
  
\bibitem{Tomalak:2014dja} 
  O.~Tomalak and M.~Vanderhaeghen,
  %``Two-photon exchange corrections in elastic muon-proton scattering,''
  Phys.\ Rev.\ D {\bf 90},  013006 (2014)
  %doi:10.1103/PhysRevD.90.013006
  [arXiv:1405.1600 [hep-ph]].
  %%CITATION = doi:10.1103/PhysRevD.90.013006;%%  
  
\bibitem{Tomalak:2015hva} 
  O.~Tomalak and M.~Vanderhaeghen,
  %``Two-photon exchange correction to muonÐproton elastic scattering at low momentum transfer,''
  Eur.\ Phys.\ J.\ C {\bf 76},  125 (2016)
  %doi:10.1140/epjc/s10052-016-3966-3
  [arXiv:1512.09113 [hep-ph]].
  %%CITATION = doi:10.1140/epjc/s10052-016-3966-3;%%  
  
\bibitem{Tomalak:2017owk} 
  O.~Tomalak,
  %``Forward two-photon exchange in elastic leptonÐproton scattering and hyperfine-splitting correction,''
  Eur.\ Phys.\ J.\ C {\bf 77}, 517 (2017)
  %doi:10.1140/epjc/s10052-017-5087-z
  [arXiv:1701.05514 [hep-ph]].
  %%CITATION = doi:10.1140/epjc/s10052-017-5087-z;%%    
  
\bibitem{Tomalak:2018jak} 
  O.~Tomalak and M.~Vanderhaeghen,
  %``Dispersion relation formalism for the two-photon exchange correction to elastic muonÐproton scattering: elastic intermediate state,''
  Eur.\ Phys.\ J.\ C {\bf 78},  514 (2018)
  %doi:10.1140/epjc/s10052-018-5988-5
  [arXiv:1803.05349 [hep-ph]].
  %%CITATION = doi:10.1140/epjc/s10052-018-5988-5;%%    

\bibitem{Pineda:2002as} 
  A.~Pineda,
  %``Leading chiral logs to the hyperfine splitting of the hydrogen and muonic hydrogen,''
  Phys.\ Rev.\ C {\bf 67}, 025201 (2003)
  %doi:10.1103/PhysRevC.67.025201
  [hep-ph/0210210].
  %%CITATION = doi:10.1103/PhysRevC.67.025201;%%  
  
\bibitem{Pineda:2004mx} 
  A.~Pineda,
  %``The Chiral structure of the Lamb shift and the definition of the proton radius,''
  Phys.\ Rev.\ C {\bf 71}, 065205 (2005)
  %doi:10.1103/PhysRevC.71.065205
  [hep-ph/0412142].
  %%CITATION = doi:10.1103/PhysRevC.71.065205;%%  
  
\bibitem{Hill:2012rh} 
  R.~J.~Hill, G.~Lee, G.~Paz and M.~P.~Solon,
  %``NRQED Lagrangian at order $1/M^4$,''
  Phys.\ Rev.\ D {\bf 87}, 053017 (2013)
  %doi:10.1103/PhysRevD.87.053017
  [arXiv:1212.4508 [hep-ph]].
  %%CITATION = doi:10.1103/PhysRevD.87.053017;%%  
  

\bibitem{Dye:2016uep} 
  S.~P.~Dye, M.~Gonderinger and G.~Paz,
  %``Elements of QED-NRQED effective field theory: NLO scattering at leading power,''
  Phys.\ Rev.\ D {\bf 94},  013006 (2016)
  %doi:10.1103/PhysRevD.94.013006
  [arXiv:1602.07770 [hep-ph]].
  %%CITATION = doi:10.1103/PhysRevD.94.013006;%%
  
  
\bibitem{Rosenbluth:1950yq} 
  M.~N.~Rosenbluth,
  %``High Energy Elastic Scattering of Electrons on Protons,''
  Phys.\ Rev.\  {\bf 79}, 615 (1950).
  %doi:10.1103/PhysRev.79.615
  %%CITATION = doi:10.1103/PhysRev.79.615;%%  
%\bibitem{Chambers:1956zz} 
E.~E.~Chambers and R.~Hofstadter,
  %``Structure of the Proton,''
  Phys.\ Rev.\  {\bf 103}, 1454 (1956).
 % doi:10.1103/PhysRev.103.1454
  %%CITATION = doi:10.1103/PhysRev.103.1454;%%  
  
  
\bibitem{Dalitz:1951ah} 
  R.~H.~Dalitz,
  %``On higher Born approximations in potential scattering,''
  Proc.\ Roy.\ Soc.\ Lond.\ A {\bf 206}, 509 (1951).
  %%CITATION = doi:10.1098/rspa.1951.0085;%%
  
\bibitem{Itzykson:1980rh} 
  C.~Itzykson and J.~B.~Zuber,
  ``Quantum Field Theory,''
  New York, Usa: Mcgraw-hill (1980) 
  
\bibitem{Dye:2018rgg} 
  S.~P.~Dye, M.~Gonderinger and G.~Paz,
  %``Elements of QED-NRQED Effective Field Theory: II. Matching of Contact Interactions,''
  Phys.\ Rev.\ D {\bf 100}, 054010 (2019)
  %doi:10.1103/PhysRevD.100.054010
  [arXiv:1812.05056 [hep-ph]].


\end{thebibliography}
\end{document}